\documentclass[prx,twocolumn, superscriptaddress]{revtex4-1}

\usepackage{epsfig}
\usepackage{graphicx}   % Include figure files
\usepackage{dcolumn}    % Align table columns on decimal point
\usepackage{bm}         % bold math
\usepackage{color}
\usepackage{amssymb,amsmath}

\usepackage{subcaption}
\usepackage{booktabs}

\begin{document}

\title{$\rm Pb_9Cu(PO4)_6O$ is a charge-transfer semiconductor}

\author{Lorenzo Celiberti}
\affiliation{University of Vienna, Faculty of Physics and Center for Computational Materials Science, Vienna, Austria}

\author{Lorenzo Varrassi}
\affiliation{Department of Physics and Astronomy 'Augusto Righi', Alma Mater Studiorum - Universit\`{a} di Bologna, Bologna, 40127 Italy}

\author{Cesare Franchini}
\affiliation{University of Vienna, Faculty of Physics and Center for Computational Materials Science, Vienna, Austria}
\affiliation{Department of Physics and Astronomy 'Augusto Righi', Alma Mater Studiorum - Universit\`{a} di Bologna, Bologna, 40127 Italy}

\begin{abstract}
By means of density functional theory and constrained random phase approximation we analyze the bandstructure of $\rm Pb_9Cu(PO4)_6O$ (named LK-99). Our data show that the lead-phosphate apatite LK-99 in the proposed Cu-doped structure is a semiconductor with predominant charge-transfer nature, a result incompatible with a superconducting behaviour. 
In order to understand the interesting electronic and magnetic properties of this compound, it will be necessary to study the actual response to doping, the possibility of alternative structural and stoichiometric (dis)orders and to clarify the magnetic interactions as well as their impact on the electronic structure.  
\end{abstract}

\maketitle

{\emph{Introduction}}
The announcement of the surprising discovery of superconductivity at room temperature and ambient pressure in  $\rm Pb_{10-x}Cu_x(PO4)_6O$ in the doping regime $0.9<x<1.1$~\cite{lee2023superconductor}
ignited an intense debate in the condensed matter community to verify the alleged superconducting nature of LK-99 and characterising its fundamental electronic properties~\cite{hou2023observation,wu2023successful,abramian2023remarks,griffin2023origin,LAI2023,si2023electronic,kurleto2023pbapatite,kumar2023absence,guo2023ferromagnetic,
zhang2023structural,yue2023correlated,bai2023ferromagnetic,korotin2023electronic,jiang2023pb9cupo46oh2,sun2023metallization,liu2023semiconducting}.
The data and analyses currently available appear to refute the interpretations initially proposed, particularly with regard to the apparent vanishing of resistivity
and the actual presence of the Meissner effect, and indicate that LK-99 is not a superconductor, either at room temperature or otherwise~\cite{kumar2023absence,guo2023ferromagnetic,liu2023semiconducting,timokhin2023synthesis,science}. Furthermore, although the synthesis procedure appears to be easily reproducible~\cite{kumar2023synthesis,wu2023successful}, doubts remain as to the precise composition, the possibility of coexistence of different phases or superstructures, and the presence of a certain degree of disorder.~\cite{krivovichev2023crystal,jiang2023pb9cupo46oh2}

\begin{figure}[h!]
\begin{subfigure}{0.9\linewidth}
   \includegraphics[ width=\linewidth]{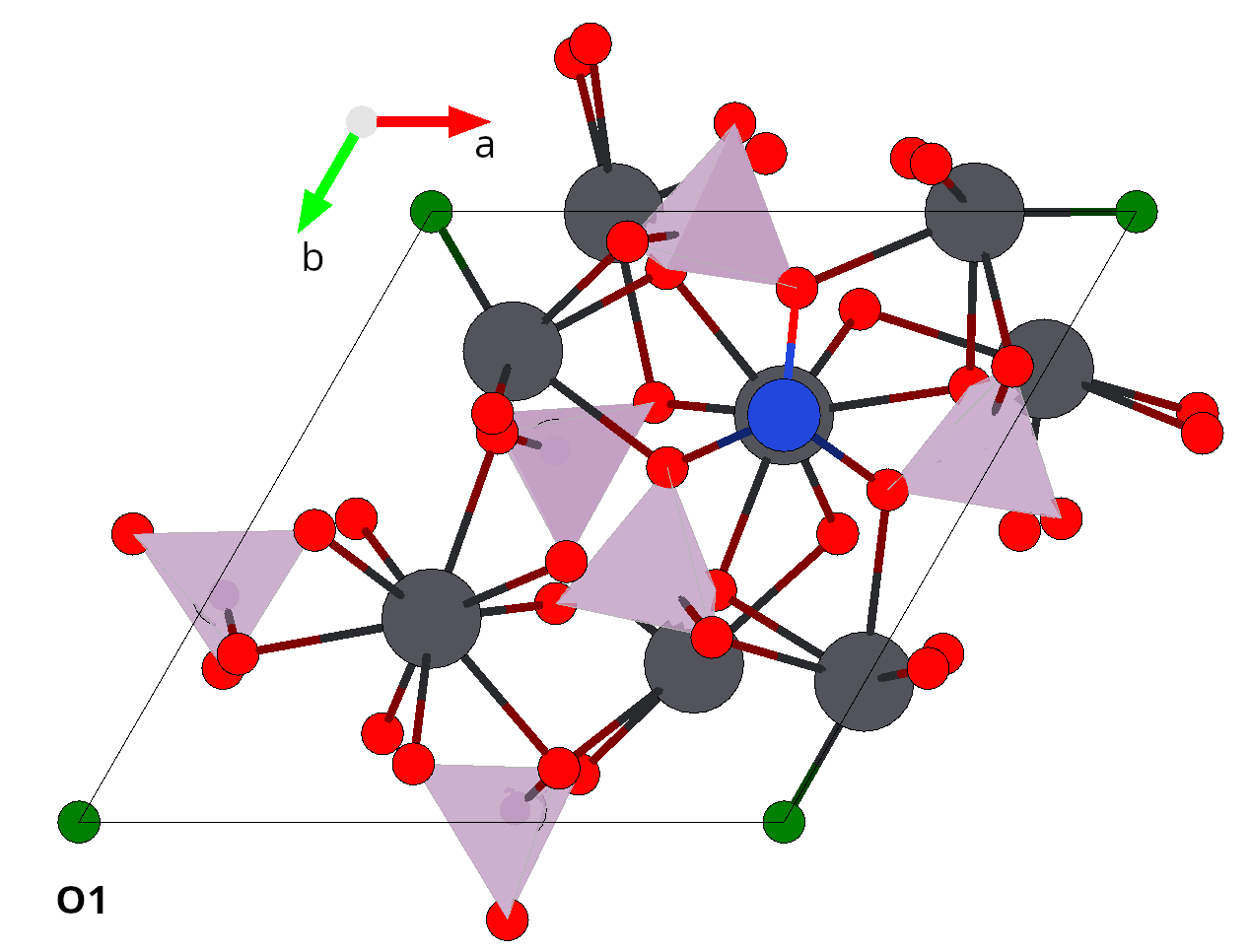}
\end{subfigure}
\vfill
\begin{subfigure}{0.99\linewidth}
   \includegraphics[width=\linewidth]{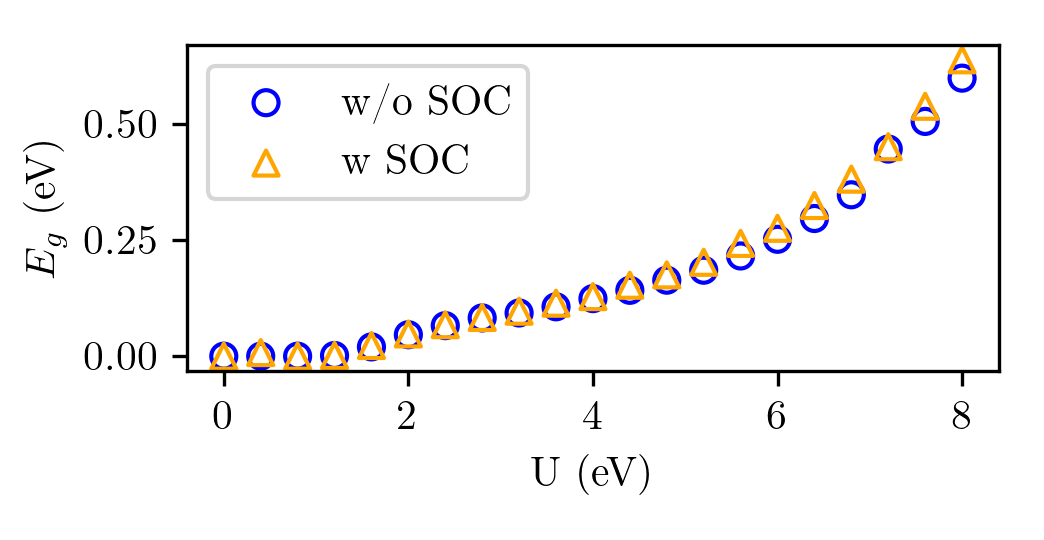}
\end{subfigure}
\caption{(a) Graphical representation of the crystal structure of $\rm Pb_9Cu(PO4)_6O$ adopted from Refs.~\cite{LAI2023} and~\cite{si2023electronic}. The shadowed tetrahedra represent the PO4 units. Large spheres indicate Pb atoms, with the blue one  showing the substitutional Cu site. Small spheres are oxygen atoms. The extra Oxygen, disconnected from the PO4 tetrahedra, is displayed in black and labeled O1. (b) Bandgap $E_g$ of $\rm Pb_{9}Cu(PO4)_6O$ as obtained from PBE+U as a function of the on-site $U$ applied to the CU-$d$ states
The PBE+U data are obtained within the complex DFT+U implementation with (w) and without (w/o) SOC. 
}\label{fig:1}
\end{figure}

\begin{figure*}[t]
%    \centering
    \includegraphics[width=\linewidth]{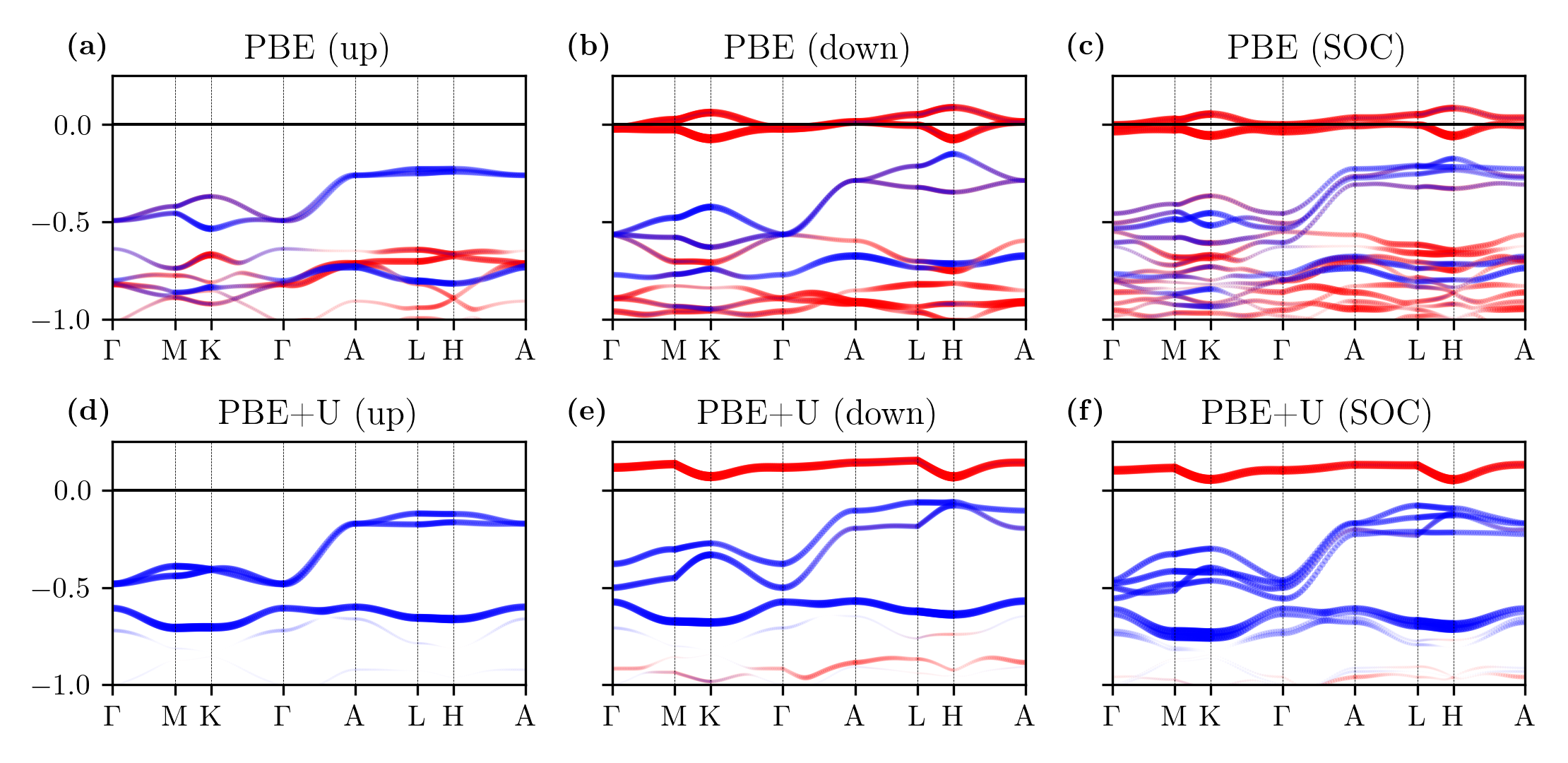}
\caption{Bandstructure of FM ordered $\rm Pb_{9}Cu(PO4)_6O$ at \textbf{(a-c)} PBE level (without U) and \textbf{(d-f)} PBE+U with $U=4.5$~eV and $J=0.8$~eV. 
For calculations non including SOC, the bands are decomposed over both spins components (up and down).
The color code identify the orbital character of the bands: Cu-d orbitals are displayed in red whereas O1-p orbitals are given in blue.}
\label{fig:2}
\end{figure*} 

\begin{figure*}[t]
%\centering
    \includegraphics[width=\linewidth]{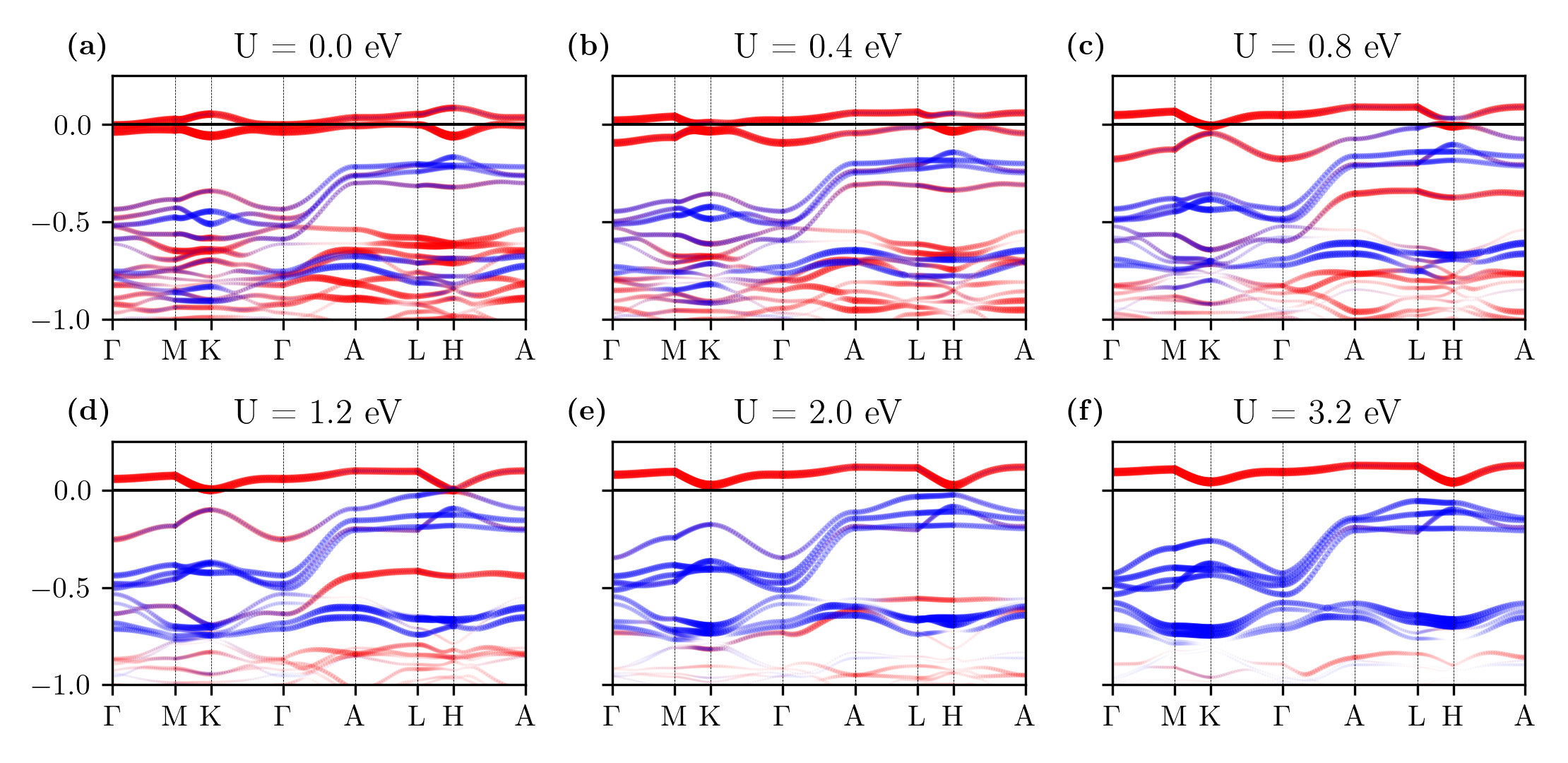}
    \caption{Bandstructure of $\rm Pb_{9}Cu(PO4)_6O$ at the PBE+U level including SOC for different values of $U$. The color code identify the orbital character of the bands: Cu-d orbitals are displayed in red whereas O1-p orbitals are given in blu. Through panels the transition from metallic \textbf{(a-d)} to CT-insulator \textbf{(e-f)} is observed.}
    \label{fig:3}
\end{figure*}

Numerous computational simulations conducted on nearly regular crystal structures with the aim of defining and elucidating the band structure have yielded only partially consistent results, in particular on the metal or semiconductor character of the $x=1$ phase $\rm Pb_{9}Cu(PO4)_6O$. 
The first Density Functional Theory (DFT) results found a metallic bandstructure, apparently robust against the application of sizable on-site Hubbard U within the DFT+U approach~\cite{LAI2023,griffin2023origin,si2023electronic,kurleto2023pbapatite,cabezasescares2023theoretical}. Alternative DFT+U calculations have instead reported the opening of gap for a sound $U$ of 4~eV~\cite{zhang2023structural}, in agreement with DFT + dynamical mean field theory (DMFT) data~\cite{yue2023correlated,korotin2023electronic}, suggesting that fractional deviation from the $x=1$ doping limit would be necessary to establish a metallic state.~\cite{yue2023correlated,si2023pb10xcuxpo46o}.
Further studies have linked the opening of the gap at $x=1$ with the inclusion of spin-orbit coupling (SOC)~\cite{bai2023ferromagnetic} or with a specific choice of the substitutional doping site~\cite{sun2023metallization}.

The aim of this work is to clarify the nature of the electronic ground state of $\rm Pb_{9}Cu(PO4)_6O$ using real and complex DFT+U methods complemented with constrained DFT calculations to assess the impact of the on-site U and SOC on the bandstructure. 
Our data indicate that using the structure used in the literature, LK-99 at $x=1$ is a semiconductor for realistic values of the on-site (static) U, above a critical $U_c \approx1.6$~eV. SOC does not play a significant role in the gap opening and only marginally modifies the band topology. The few hundred meV large gap is predominantly charge-transfer in nature for U values in the regime predicted by the constrained random phase approximation (cRPA) ($\approx$ 3.5-5.5 eV, in agreement with previous reports~\cite{yue2023correlated}).

{\emph{Computational details}}
We used the Vienna {\emph{ab initio}} simulation Package (VASP)~\cite{PhysRevB.47.558,PhysRevB.54.11169} within DFT in the Perdew-Burke-Ernzerhof 9PBE) exchange-correlation functional. We have tested two different schemes for $+U$ corrections, the Dudarev~\cite{PhysRevB.57.1505} and the Lichtenstein~\cite{PhysRevB.52.R5467} approaches, which have been proved to deliver the same outcome. 
For both DFT+U schemes we applied two different implementations: the conventional real version ({\tt{std}}, in VASP jargon) and the complex representation ({\tt{ncl}}, in VASP jargon). 
The effect of  SOC has been inspected within complex DFT+U. 
We used an energy cutoff of $600$~eV 
and a k-points grid of $6\times 6\times 8$.
All calculations were performed on the crystal structure reported in literature~\cite{griffin2023origin,LAI2023,si2023electronic}, assuming a ferromagnetic ordering (FM).
cRPA calculations were performed using VASP optimized projected localized orbitals (PLOs)~\cite{karolak_general_2011,S2018,amadon2008} for constructing the correlated subspace.

{\emph{Results and discussions}}
The parent compound of LK-99, lead-phosphate apatite with chemical formula $\rm Pb_{10}(PO4)_6O$~\cite{Merker,krivovichev2023crystal} is a nonmagnetic insulator with a DFT bandgap of about 2.7 eV opened between the O-p and Pb-p states eV~\cite{LAI2023,si2023electronic}.
Replacing Pb with Cu causes the formation of low-dispersing correlated Cu $d$-bands in the Fermi region~\cite{griffin2023origin,LAI2023,si2023electronic}. 
Using real DFT+U we obtain a metallic solution regardless on the value of $U$, confirming previous reports~\cite{LAI2023,si2023electronic}.
However, within complex DFT+U, which enables the breaking of orbital symmetry in the Cu-$d$ states, the situation changes, as shown in Figs.~\ref{fig:1}. 
We find that the electronic ground state is highly sensible on the value of $U$.
In PBE, e.g. neglecting corrections to improve the treatment of electronic correlation, LK-99 exhibits a metallic ground state, characterized by a pair of entangled Cu-$d$ bands crossing the Fermi energy (see Fig.\ref{fig:2}). 
However, including an on-site Hubbard $U$ larger than 1.6~eV 
causes the opening of a small gap which increases as a function of $U$. 
Enabling SOC does not modify this result, although in the $U=0$ (PBE) limit SOC splits the pair of correlated Cu-$d$ near Fermi establishing a weak metallic state~\cite{bai2023ferromagnetic}. 
The insulating phase obtained by complex DFT+U is energetically more favorable than the corresponding real DFT+U result and represents ground state of $\rm Pb_9Cu(PO4)_6O$. We believe that, a similar gapped phase could be obtained by breaking the orbital symmetry in other ways, for instance by suitably initializing the orbital occupation of the Cu-$d$ shell, by constructing distorted supercell structures or by a imposing non-collinear spin orientation as starting point.

To quantify the degree of electronic correlation in $\rm Pb_{9}Cu(PO4)_6O$ we have extracted the effective on-site Coulomb (U) and exchange (J) interactions for the correlated Cu-$d$ bands from cRPA adopting two different low-energy effective models depending on the numbers of states included in the correlated basis and those included in the screening channel.
In the first one ($d-d$), we used Cu-$d_{xz}$ and $d_{yz}$ projectors and excluded from the screening the two topmost bands in Fig.~\ref{fig:2}(b), which have mostly $d$-character.
In the second model ($p+d$), O1-$p_x$ and $p_y$ projectors are also considered in the construction of the correlated subspace and the four topmost bands are excluded from screening.

\begin{table}[ht]
\begin{tabular}{@{}cccccc@{}}
  &   & \multicolumn{2}{c}{this work} & \multicolumn{2}{c}{Yue~\emph{et al.}~\cite{yue2023correlated}} \\ \midrule
model & n. of bands  &U (eV)        & J (eV)        & U (eV)       & J (eV)      \\ \midrule
$d-d$ & 2 &3.53          & 0.77          & 2.94         & 0.61        \\
%d-pd & &              &               & -            & -           \\
$p+d$ & 4 &5.51          & 0.78          & 5.67         & 0.65        \\ \bottomrule
\end{tabular}
\caption{Interactions parameters $U$ anad $J$ obtained by cRPA, compared with the values of Yue~\emph{et al.}~\cite{yue2023correlated}.}
\label{tab:cRPA}
\end{table}

The interaction parameters obtained, collected in Tab.~\ref{tab:cRPA}, indicate that $U$ and $J$ are to a good extent similar to those of cuprates~\cite{PhysRevB.91.125142} as noted in previous studies~\cite{yue2023correlated,si2023pb10xcuxpo46o}, and in good agreement with the cRPA data of Yue {\emph et al.}~\cite{yue2023correlated}. We obtained $U_{dd}$ between 3.5~eV ($d-d$ model) and 5.5~eV ($p+d$ model), while $J$ is about 0.8~eV, essentially insensible to the model.
The band structure obtained using the representative value U-J=3.7~eV are displayed in Fig.~\ref{fig:2}. 
Based on the above data, we conclude that our results clearly indicate that LK-99 at x=1 is a semiconductor in which the gap is open due to correlation effects, not necessarily of a dynamic nature~\cite{zhang2023structural,si2023pb10xcuxpo46o,yue2023correlated,korotin2023electronic}.

In a previous tight-binding study, Li {\emph{et al.}}~\cite{si2023pb10xcuxpo46o} have argued that LK-99 should exhibit either a Mott or a charge-transfer (CT) like nature~\cite{mit}. In a Mott insulator the  
gap is solely accounted by $d$ bands and is opened between a lower Hubbard band and upper Hubbard band~\cite{mott, hubbard}.
In a CT insulator, instead, the role of $p$ states can not be neglected and the gap is formed between an
occupied (predominantly) $p$ band and an empty $d$ band~\cite{ct}.
By inspecting the change in the orbital character in the DFT+U electronic bands near $E_F$ induced by $U$ (see Fig.~\ref{fig:2}), it can be concluded that LK-99 belongs to the class of charge-transfer insulator. Without $U$ the electronic state is represented by two entangled Cu-$d$ bands with  
$d_{yz}/d_{x^2-y^2}$ (upper) and $d_{xz}/d_{xy}$ (lower) bands  followed by lower-energy O$-p$ states. The inclusion of $U$, rather than rigidly separating the two Cu bands near $E_F$, thus yielding a Mott solution, induces a substantial transfer of $d$ electrons at lower energies leading to the formation of a $d-p$ gap that can be categorized as CT. 
The transition from the metallic $U=0$ solution to the CT state can be appreciated in Fig.~\ref{fig:3}, were we show the evolution of the low-energy bands as a function of $U$.
In the weakly correlated regime ($U<1$~eV), the two d-bands structure is roughly preserved, but for larger $U$ there is a rapid shift from the Mott picture with the Cu-$d$ electrons that are transferred to energy below -1~eV.

{\emph{Conclusion and perspective.}}
In summary, according to our simulations, single crystal LK-99 does not exhibit a metallic behaviour. The improved treatment of electronic correlation through realistic values of on-site $U$ and enabling the breaking of orbital symmetry within the Cu-$d$ states,
opens a gap of predominant CT character, 
a behaviour not compatible with superconductivity. 
There are some important aspects that will have to be studied in order to clarify the general validity of these conclusions in realistic conditions. 
Firstly, recent experiments provide evidence that the LK-99 compound may have a different stoichoimetry and structure than initially proposed~\cite{jiang2023pb9cupo46oh2,jain2023phase,thakur2023synthesis,krivovichev2023crystal} including uncertainties on the geometrical distribution of the extra oxygen. 
The second relevant issue is substitutional doping.
Cu substitution is considered highly thermodynamically disfavored~\cite{jiang2023pb9cupo46oh2} and the incorporation Cu at Pb sites proves to be extremely difficult~\cite{thakur2023synthesis}. Moreover, LK-99 synthetized with $Cu_2S$ have a significant fraction of copper sulfide, which could be responsible for the sharp transitions in electrical resistivity and heat capacity~\cite{jain2023phase}. This experimental evidence casts doubt on the possibility of achieving a realistic metallic state by doping the $x=1$ insulating bands with holes or electrons~\cite{liu2023semiconducting,yue2023correlated}: it could be achieved in idealised computational simulations, but would be difficult to realise in real samples with this structure and stoichiometry.
As a final point, it will be necessary to understand the magnetic nature of this material, which at the moment remains quite debated, in particular the diamagnetic or ferromagnetic character, the relationship with the levitation effect and the possible interference between the magnetic and electronic degree of freedom~\cite{jiang2023pb9cupo46oh2,guo2023ferromagnetic,sun2023metallization}. 

To conclude, although there is now widespread awareness that LK-99 is not the room-temperature SC that was hoped for ave, it nevertheless appears to be a material with very interesting properties not yet understood, that could provide a basis for further studies and potential applications.

\section{Acknowledgements}
CF and LC acknowledge the Austrian Science Fund (FWF) and the University of Vienna for continuous support. CF
acknowledges financial support from the Italian Ministry for Research and Education through
PRIN-2022 project 2022L28H97 (IT-MIUR Grant No. 2022L28H97). LC acknowledges the Vienna Doctoral School of Physics.
Computing time at the Vienna Scientific Cluster is greatly acknowledged.

\bibliographystyle{apsrev4-1}
\bibliography{references}

%merlin.mbs apsrev4-1.bst 2010-07-25 4.21a (PWD, AO, DPC) hacked
%Control: key (0)
%Control: author (72) initials jnrlst
%Control: editor formatted (1) identically to author
%Control: production of article title (-1) disabled
%Control: page (0) single
%Control: year (1) truncated
%Control: production of eprint (0) enabled
\begin{thebibliography}{38}%
\makeatletter
\providecommand \@ifxundefined [1]{%
 \@ifx{#1\undefined}
}%
\providecommand \@ifnum [1]{%
 \ifnum #1\expandafter \@firstoftwo
 \else \expandafter \@secondoftwo
 \fi
}%
\providecommand \@ifx [1]{%
 \ifx #1\expandafter \@firstoftwo
 \else \expandafter \@secondoftwo
 \fi
}%
\providecommand \natexlab [1]{#1}%
\providecommand \enquote  [1]{``#1''}%
\providecommand \bibnamefont  [1]{#1}%
\providecommand \bibfnamefont [1]{#1}%
\providecommand \citenamefont [1]{#1}%
\providecommand \href@noop [0]{\@secondoftwo}%
\providecommand \href [0]{\begingroup \@sanitize@url \@href}%
\providecommand \@href[1]{\@@startlink{#1}\@@href}%
\providecommand \@@href[1]{\endgroup#1\@@endlink}%
\providecommand \@sanitize@url [0]{\catcode `\\12\catcode `\$12\catcode
  `\&12\catcode `\#12\catcode `\^12\catcode `\_12\catcode `\%12\relax}%
\providecommand \@@startlink[1]{}%
\providecommand \@@endlink[0]{}%
\providecommand \url  [0]{\begingroup\@sanitize@url \@url }%
\providecommand \@url [1]{\endgroup\@href {#1}{\urlprefix }}%
\providecommand \urlprefix  [0]{URL }%
\providecommand \Eprint [0]{\href }%
\providecommand \doibase [0]{http://dx.doi.org/}%
\providecommand \selectlanguage [0]{\@gobble}%
\providecommand \bibinfo  [0]{\@secondoftwo}%
\providecommand \bibfield  [0]{\@secondoftwo}%
\providecommand \translation [1]{[#1]}%
\providecommand \BibitemOpen [0]{}%
\providecommand \bibitemStop [0]{}%
\providecommand \bibitemNoStop [0]{.\EOS\space}%
\providecommand \EOS [0]{\spacefactor3000\relax}%
\providecommand \BibitemShut  [1]{\csname bibitem#1\endcsname}%
\let\auto@bib@innerbib\@empty
%</preamble>
\bibitem [{\citenamefont {Lee}\ \emph {et~al.}(2023)\citenamefont {Lee},
  \citenamefont {Kim}, \citenamefont {Kim}, \citenamefont {Im}, \citenamefont
  {An},\ and\ \citenamefont {Auh}}]{lee2023superconductor}%
  \BibitemOpen
  \bibfield  {author} {\bibinfo {author} {\bibfnamefont {S.}~\bibnamefont
  {Lee}}, \bibinfo {author} {\bibfnamefont {J.}~\bibnamefont {Kim}}, \bibinfo
  {author} {\bibfnamefont {H.-T.}\ \bibnamefont {Kim}}, \bibinfo {author}
  {\bibfnamefont {S.}~\bibnamefont {Im}}, \bibinfo {author} {\bibfnamefont
  {S.}~\bibnamefont {An}}, \ and\ \bibinfo {author} {\bibfnamefont {K.~H.}\
  \bibnamefont {Auh}},\ }\href@noop {} {\enquote {\bibinfo {title}
  {Superconductor pb$_{10-x}$cu$_x$(po$_4$)$_6$o showing levitation at room
  temperature and atmospheric pressure and mechanism},}\ } (\bibinfo {year}
  {2023}),\ \Eprint {http://arxiv.org/abs/2307.12037} {arXiv:2307.12037
  [cond-mat.supr-con]} \BibitemShut {NoStop}%
\bibitem [{\citenamefont {Hou}\ \emph {et~al.}(2023)\citenamefont {Hou},
  \citenamefont {Wei}, \citenamefont {Zhou}, \citenamefont {Sun},\ and\
  \citenamefont {Shi}}]{hou2023observation}%
  \BibitemOpen
  \bibfield  {author} {\bibinfo {author} {\bibfnamefont {Q.}~\bibnamefont
  {Hou}}, \bibinfo {author} {\bibfnamefont {W.}~\bibnamefont {Wei}}, \bibinfo
  {author} {\bibfnamefont {X.}~\bibnamefont {Zhou}}, \bibinfo {author}
  {\bibfnamefont {Y.}~\bibnamefont {Sun}}, \ and\ \bibinfo {author}
  {\bibfnamefont {Z.}~\bibnamefont {Shi}},\ }\href@noop {} {\enquote {\bibinfo
  {title} {Observation of zero resistance above 100$^\circ$ k in
  pb$_{10-x}$cu$_x$(po$_4$)$_6$o},}\ } (\bibinfo {year} {2023}),\ \Eprint
  {http://arxiv.org/abs/2308.01192} {arXiv:2308.01192 [cond-mat.supr-con]}
  \BibitemShut {NoStop}%
\bibitem [{\citenamefont {Wu}\ \emph {et~al.}(2023)\citenamefont {Wu},
  \citenamefont {Yang}, \citenamefont {Xiao},\ and\ \citenamefont
  {Chang}}]{wu2023successful}%
  \BibitemOpen
  \bibfield  {author} {\bibinfo {author} {\bibfnamefont {H.}~\bibnamefont
  {Wu}}, \bibinfo {author} {\bibfnamefont {L.}~\bibnamefont {Yang}}, \bibinfo
  {author} {\bibfnamefont {B.}~\bibnamefont {Xiao}}, \ and\ \bibinfo {author}
  {\bibfnamefont {H.}~\bibnamefont {Chang}},\ }\href@noop {} {\enquote
  {\bibinfo {title} {Successful growth and room temperature ambient-pressure
  magnetic levitation of lk-99},}\ } (\bibinfo {year} {2023}),\ \Eprint
  {http://arxiv.org/abs/2308.01516} {arXiv:2308.01516 [cond-mat.supr-con]}
  \BibitemShut {NoStop}%
\bibitem [{\citenamefont {Abramian}\ \emph {et~al.}(2023)\citenamefont
  {Abramian}, \citenamefont {Kuzanyan}, \citenamefont {Nikoghosyan},
  \citenamefont {Teknowijoyo},\ and\ \citenamefont
  {Gulian}}]{abramian2023remarks}%
  \BibitemOpen
  \bibfield  {author} {\bibinfo {author} {\bibfnamefont {P.}~\bibnamefont
  {Abramian}}, \bibinfo {author} {\bibfnamefont {A.}~\bibnamefont {Kuzanyan}},
  \bibinfo {author} {\bibfnamefont {V.}~\bibnamefont {Nikoghosyan}}, \bibinfo
  {author} {\bibfnamefont {S.}~\bibnamefont {Teknowijoyo}}, \ and\ \bibinfo
  {author} {\bibfnamefont {A.}~\bibnamefont {Gulian}},\ }\href@noop {}
  {\enquote {\bibinfo {title} {Some remarks on possible superconductivity of
  composition pb$_9$cup$_6$o$_{25}$},}\ } (\bibinfo {year} {2023}),\ \Eprint
  {http://arxiv.org/abs/2308.01723} {arXiv:2308.01723 [cond-mat.supr-con]}
  \BibitemShut {NoStop}%
\bibitem [{\citenamefont {Griffin}(2023)}]{griffin2023origin}%
  \BibitemOpen
  \bibfield  {author} {\bibinfo {author} {\bibfnamefont {S.~M.}\ \bibnamefont
  {Griffin}},\ }\href@noop {} {\enquote {\bibinfo {title} {Origin of correlated
  isolated flat bands in copper-substituted lead phosphate apatite},}\ }
  (\bibinfo {year} {2023}),\ \Eprint {http://arxiv.org/abs/2307.16892}
  {arXiv:2307.16892 [cond-mat.supr-con]} \BibitemShut {NoStop}%
\bibitem [{\citenamefont {Lai}\ \emph {et~al.}(2023)\citenamefont {Lai},
  \citenamefont {Li}, \citenamefont {Liu}, \citenamefont {Sun},\ and\
  \citenamefont {Chen}}]{LAI2023}%
  \BibitemOpen
  \bibfield  {author} {\bibinfo {author} {\bibfnamefont {J.}~\bibnamefont
  {Lai}}, \bibinfo {author} {\bibfnamefont {J.}~\bibnamefont {Li}}, \bibinfo
  {author} {\bibfnamefont {P.}~\bibnamefont {Liu}}, \bibinfo {author}
  {\bibfnamefont {Y.}~\bibnamefont {Sun}}, \ and\ \bibinfo {author}
  {\bibfnamefont {X.-Q.}\ \bibnamefont {Chen}},\ }\href {\doibase
  https://doi.org/10.1016/j.jmst.2023.08.001} {\bibfield  {journal} {\bibinfo
  {journal} {{Journal of Materials Science \& Technology}}\ } (\bibinfo {year}
  {2023}),\ https://doi.org/10.1016/j.jmst.2023.08.001}\BibitemShut {NoStop}%
\bibitem [{\citenamefont {Si}\ and\ \citenamefont
  {Held}(2023)}]{si2023electronic}%
  \BibitemOpen
  \bibfield  {author} {\bibinfo {author} {\bibfnamefont {L.}~\bibnamefont
  {Si}}\ and\ \bibinfo {author} {\bibfnamefont {K.}~\bibnamefont {Held}},\
  }\href@noop {} {\enquote {\bibinfo {title} {Electronic structure of the
  putative room-temperature superconductor $\mathrm{Pb_9Cu(PO_4)_6O}$},}\ }
  (\bibinfo {year} {2023}),\ \Eprint {http://arxiv.org/abs/2308.00676}
  {arXiv:2308.00676 [cond-mat.supr-con]} \BibitemShut {NoStop}%
\bibitem [{\citenamefont {Kurleto}\ \emph {et~al.}(2023)\citenamefont
  {Kurleto}, \citenamefont {Lany}, \citenamefont {Pashov}, \citenamefont
  {Acharya}, \citenamefont {van Schilfgaarde},\ and\ \citenamefont
  {Dessau}}]{kurleto2023pbapatite}%
  \BibitemOpen
  \bibfield  {author} {\bibinfo {author} {\bibfnamefont {R.}~\bibnamefont
  {Kurleto}}, \bibinfo {author} {\bibfnamefont {S.}~\bibnamefont {Lany}},
  \bibinfo {author} {\bibfnamefont {D.}~\bibnamefont {Pashov}}, \bibinfo
  {author} {\bibfnamefont {S.}~\bibnamefont {Acharya}}, \bibinfo {author}
  {\bibfnamefont {M.}~\bibnamefont {van Schilfgaarde}}, \ and\ \bibinfo
  {author} {\bibfnamefont {D.~S.}\ \bibnamefont {Dessau}},\ }\href@noop {}
  {\enquote {\bibinfo {title} {Pb-apatite framework as a generator of novel
  flat-band cuo based physics, including possible room temperature
  superconductivity},}\ } (\bibinfo {year} {2023}),\ \Eprint
  {http://arxiv.org/abs/2308.00698} {arXiv:2308.00698 [cond-mat.supr-con]}
  \BibitemShut {NoStop}%
\bibitem [{\citenamefont {Kumar}\ \emph
  {et~al.}(2023{\natexlab{a}})\citenamefont {Kumar}, \citenamefont {Karn},
  \citenamefont {Kumar},\ and\ \citenamefont {Awana}}]{kumar2023absence}%
  \BibitemOpen
  \bibfield  {author} {\bibinfo {author} {\bibfnamefont {K.}~\bibnamefont
  {Kumar}}, \bibinfo {author} {\bibfnamefont {N.~K.}\ \bibnamefont {Karn}},
  \bibinfo {author} {\bibfnamefont {Y.}~\bibnamefont {Kumar}}, \ and\ \bibinfo
  {author} {\bibfnamefont {V.~P.~S.}\ \bibnamefont {Awana}},\ }\href@noop {}
  {\enquote {\bibinfo {title} {Absence of superconductivity in lk-99 at ambient
  conditions},}\ } (\bibinfo {year} {2023}{\natexlab{a}}),\ \Eprint
  {http://arxiv.org/abs/2308.03544} {arXiv:2308.03544 [cond-mat.supr-con]}
  \BibitemShut {NoStop}%
\bibitem [{\citenamefont {Guo}\ \emph {et~al.}(2023)\citenamefont {Guo},
  \citenamefont {Li},\ and\ \citenamefont {Jia}}]{guo2023ferromagnetic}%
  \BibitemOpen
  \bibfield  {author} {\bibinfo {author} {\bibfnamefont {K.}~\bibnamefont
  {Guo}}, \bibinfo {author} {\bibfnamefont {Y.}~\bibnamefont {Li}}, \ and\
  \bibinfo {author} {\bibfnamefont {S.}~\bibnamefont {Jia}},\ }\href@noop {}
  {\enquote {\bibinfo {title} {Ferromagnetic half levitation of lk-99-like
  synthetic samples},}\ } (\bibinfo {year} {2023}),\ \Eprint
  {http://arxiv.org/abs/2308.03110} {arXiv:2308.03110 [cond-mat.supr-con]}
  \BibitemShut {NoStop}%
\bibitem [{\citenamefont {Zhang}\ \emph {et~al.}(2023)\citenamefont {Zhang},
  \citenamefont {Li}, \citenamefont {Yan}, \citenamefont {Gao}, \citenamefont
  {Ma}, \citenamefont {Yan},\ and\ \citenamefont {Xie}}]{zhang2023structural}%
  \BibitemOpen
  \bibfield  {author} {\bibinfo {author} {\bibfnamefont {J.}~\bibnamefont
  {Zhang}}, \bibinfo {author} {\bibfnamefont {H.}~\bibnamefont {Li}}, \bibinfo
  {author} {\bibfnamefont {M.}~\bibnamefont {Yan}}, \bibinfo {author}
  {\bibfnamefont {M.}~\bibnamefont {Gao}}, \bibinfo {author} {\bibfnamefont
  {F.}~\bibnamefont {Ma}}, \bibinfo {author} {\bibfnamefont {X.-W.}\
  \bibnamefont {Yan}}, \ and\ \bibinfo {author} {\bibfnamefont {Z.~Y.}\
  \bibnamefont {Xie}},\ }\href@noop {} {\enquote {\bibinfo {title} {Structural,
  electronic, magnetic properties of cu-doped lead-apatite
  $\mathrm{Pb_{10-x}Cu_x(PO_4)_6O}$},}\ } (\bibinfo {year} {2023}),\ \Eprint
  {http://arxiv.org/abs/2308.04344} {arXiv:2308.04344 [cond-mat.supr-con]}
  \BibitemShut {NoStop}%
\bibitem [{\citenamefont {Yue}\ \emph {et~al.}(2023)\citenamefont {Yue},
  \citenamefont {Christiansson},\ and\ \citenamefont
  {Werner}}]{yue2023correlated}%
  \BibitemOpen
  \bibfield  {author} {\bibinfo {author} {\bibfnamefont {C.}~\bibnamefont
  {Yue}}, \bibinfo {author} {\bibfnamefont {V.}~\bibnamefont {Christiansson}},
  \ and\ \bibinfo {author} {\bibfnamefont {P.}~\bibnamefont {Werner}},\
  }\href@noop {} {\enquote {\bibinfo {title} {Correlated electronic structure
  of $\mathrm{Pb_{10-x}Cu_x(PO_4)_6O}$},}\ } (\bibinfo {year} {2023}),\ \Eprint
  {http://arxiv.org/abs/2308.04976} {arXiv:2308.04976 [cond-mat.str-el]}
  \BibitemShut {NoStop}%
\bibitem [{\citenamefont {Bai}\ \emph {et~al.}(2023)\citenamefont {Bai},
  \citenamefont {Gao},\ and\ \citenamefont {Zeng}}]{bai2023ferromagnetic}%
  \BibitemOpen
  \bibfield  {author} {\bibinfo {author} {\bibfnamefont {H.}~\bibnamefont
  {Bai}}, \bibinfo {author} {\bibfnamefont {L.}~\bibnamefont {Gao}}, \ and\
  \bibinfo {author} {\bibfnamefont {C.}~\bibnamefont {Zeng}},\ }\href@noop {}
  {\enquote {\bibinfo {title} {Ferromagnetic ground state and spin-orbit
  coupling induced bandgap open in lk99},}\ } (\bibinfo {year} {2023}),\
  \Eprint {http://arxiv.org/abs/2308.05134} {arXiv:2308.05134
  [cond-mat.supr-con]} \BibitemShut {NoStop}%
\bibitem [{\citenamefont {Korotin}\ \emph {et~al.}(2023)\citenamefont
  {Korotin}, \citenamefont {Novoselov}, \citenamefont {Shorikov}, \citenamefont
  {Anisimov},\ and\ \citenamefont {Oganov}}]{korotin2023electronic}%
  \BibitemOpen
  \bibfield  {author} {\bibinfo {author} {\bibfnamefont {D.~M.}\ \bibnamefont
  {Korotin}}, \bibinfo {author} {\bibfnamefont {D.~Y.}\ \bibnamefont
  {Novoselov}}, \bibinfo {author} {\bibfnamefont {A.~O.}\ \bibnamefont
  {Shorikov}}, \bibinfo {author} {\bibfnamefont {V.~I.}\ \bibnamefont
  {Anisimov}}, \ and\ \bibinfo {author} {\bibfnamefont {A.~R.}\ \bibnamefont
  {Oganov}},\ }\href@noop {} {\enquote {\bibinfo {title} {Electronic
  correlations in promising room-temperature superconductor
  pb$_9$cu(po$_4$)$_6$o: a dft+dmft study},}\ } (\bibinfo {year} {2023}),\
  \Eprint {http://arxiv.org/abs/2308.04301} {arXiv:2308.04301
  [cond-mat.supr-con]} \BibitemShut {NoStop}%
\bibitem [{\citenamefont {Jiang}\ \emph {et~al.}(2023)\citenamefont {Jiang},
  \citenamefont {Lee}, \citenamefont {Herzog-Arbeitman}, \citenamefont {Yu},
  \citenamefont {Feng}, \citenamefont {Hu}, \citenamefont {Călugăru},
  \citenamefont {Brodale}, \citenamefont {Gormley}, \citenamefont {Vergniory},
  \citenamefont {Felser}, \citenamefont {Blanco-Canosa}, \citenamefont
  {Hendon}, \citenamefont {Schoop},\ and\ \citenamefont
  {Bernevig}}]{jiang2023pb9cupo46oh2}%
  \BibitemOpen
  \bibfield  {author} {\bibinfo {author} {\bibfnamefont {Y.}~\bibnamefont
  {Jiang}}, \bibinfo {author} {\bibfnamefont {S.~B.}\ \bibnamefont {Lee}},
  \bibinfo {author} {\bibfnamefont {J.}~\bibnamefont {Herzog-Arbeitman}},
  \bibinfo {author} {\bibfnamefont {J.}~\bibnamefont {Yu}}, \bibinfo {author}
  {\bibfnamefont {X.}~\bibnamefont {Feng}}, \bibinfo {author} {\bibfnamefont
  {H.}~\bibnamefont {Hu}}, \bibinfo {author} {\bibfnamefont {D.}~\bibnamefont
  {Călugăru}}, \bibinfo {author} {\bibfnamefont {P.~S.}\ \bibnamefont
  {Brodale}}, \bibinfo {author} {\bibfnamefont {E.~L.}\ \bibnamefont
  {Gormley}}, \bibinfo {author} {\bibfnamefont {M.~G.}\ \bibnamefont
  {Vergniory}}, \bibinfo {author} {\bibfnamefont {C.}~\bibnamefont {Felser}},
  \bibinfo {author} {\bibfnamefont {S.}~\bibnamefont {Blanco-Canosa}}, \bibinfo
  {author} {\bibfnamefont {C.~H.}\ \bibnamefont {Hendon}}, \bibinfo {author}
  {\bibfnamefont {L.~M.}\ \bibnamefont {Schoop}}, \ and\ \bibinfo {author}
  {\bibfnamefont {B.~A.}\ \bibnamefont {Bernevig}},\ }\href@noop {} {\enquote
  {\bibinfo {title} {Pb$_9$cu(po4)$_6$(oh)$_2$: Phonon bands, localized flat
  band magnetism, models, and chemical analysis},}\ } (\bibinfo {year}
  {2023}),\ \Eprint {http://arxiv.org/abs/2308.05143} {arXiv:2308.05143
  [cond-mat.supr-con]} \BibitemShut {NoStop}%
\bibitem [{\citenamefont {Sun}\ \emph {et~al.}(2023)\citenamefont {Sun},
  \citenamefont {Ho},\ and\ \citenamefont {Antropov}}]{sun2023metallization}%
  \BibitemOpen
  \bibfield  {author} {\bibinfo {author} {\bibfnamefont {Y.}~\bibnamefont
  {Sun}}, \bibinfo {author} {\bibfnamefont {K.-M.}\ \bibnamefont {Ho}}, \ and\
  \bibinfo {author} {\bibfnamefont {V.}~\bibnamefont {Antropov}},\ }\href@noop
  {} {\enquote {\bibinfo {title} {Metallization and spin fluctuations in
  cu-doped lead apatite},}\ } (\bibinfo {year} {2023}),\ \Eprint
  {http://arxiv.org/abs/2308.03454} {arXiv:2308.03454 [cond-mat.supr-con]}
  \BibitemShut {NoStop}%
\bibitem [{\citenamefont {Liu}\ \emph {et~al.}(2023)\citenamefont {Liu},
  \citenamefont {Meng}, \citenamefont {Wang}, \citenamefont {Chen},
  \citenamefont {Duan}, \citenamefont {Zhou}, \citenamefont {Yan},
  \citenamefont {Qin},\ and\ \citenamefont {Liu}}]{liu2023semiconducting}%
  \BibitemOpen
  \bibfield  {author} {\bibinfo {author} {\bibfnamefont {L.}~\bibnamefont
  {Liu}}, \bibinfo {author} {\bibfnamefont {Z.}~\bibnamefont {Meng}}, \bibinfo
  {author} {\bibfnamefont {X.}~\bibnamefont {Wang}}, \bibinfo {author}
  {\bibfnamefont {H.}~\bibnamefont {Chen}}, \bibinfo {author} {\bibfnamefont
  {Z.}~\bibnamefont {Duan}}, \bibinfo {author} {\bibfnamefont {X.}~\bibnamefont
  {Zhou}}, \bibinfo {author} {\bibfnamefont {H.}~\bibnamefont {Yan}}, \bibinfo
  {author} {\bibfnamefont {P.}~\bibnamefont {Qin}}, \ and\ \bibinfo {author}
  {\bibfnamefont {Z.}~\bibnamefont {Liu}},\ }\href@noop {} {\enquote {\bibinfo
  {title} {Semiconducting transport in pb$_{10-x}$cu$_x$(po$_4$)$_6$o sintered
  from pb$_2$so$_5$ and cu$_3$p},}\ } (\bibinfo {year} {2023}),\ \Eprint
  {http://arxiv.org/abs/2307.16802} {arXiv:2307.16802 [cond-mat.supr-con]}
  \BibitemShut {NoStop}%
\bibitem [{\citenamefont {Timokhin}\ \emph {et~al.}(2023)\citenamefont
  {Timokhin}, \citenamefont {Chen}, \citenamefont {Yang},\ and\ \citenamefont
  {Mishchenko}}]{timokhin2023synthesis}%
  \BibitemOpen
  \bibfield  {author} {\bibinfo {author} {\bibfnamefont {I.}~\bibnamefont
  {Timokhin}}, \bibinfo {author} {\bibfnamefont {C.}~\bibnamefont {Chen}},
  \bibinfo {author} {\bibfnamefont {Q.}~\bibnamefont {Yang}}, \ and\ \bibinfo
  {author} {\bibfnamefont {A.}~\bibnamefont {Mishchenko}},\ }\href@noop {}
  {\enquote {\bibinfo {title} {Synthesis and characterisation of lk-99},}\ }
  (\bibinfo {year} {2023}),\ \Eprint {http://arxiv.org/abs/2308.03823}
  {arXiv:2308.03823 [cond-mat.supr-con]} \BibitemShut {NoStop}%
\bibitem [{\citenamefont {Cho}(2023)}]{science}%
  \BibitemOpen
  \bibfield  {author} {\bibinfo {author} {\bibfnamefont {A.}~\bibnamefont
  {Cho}},\ }\href@noop {} {\enquote {\bibinfo {title} {The short, spectacular
  life of that viral room-temperature superconductivity claim},}\ }\bibinfo
  {howpublished}
  {\url{https://www.science.org/content/article/short-spectacular-life-viral-room-temperature-superconductivity-claim}}
  (\bibinfo {year} {2023}),\ \bibinfo {note} {accessed: 2023-08-12}\BibitemShut
  {NoStop}%
\bibitem [{\citenamefont {Kumar}\ \emph
  {et~al.}(2023{\natexlab{b}})\citenamefont {Kumar}, \citenamefont {Karn},\
  and\ \citenamefont {Awana}}]{kumar2023synthesis}%
  \BibitemOpen
  \bibfield  {author} {\bibinfo {author} {\bibfnamefont {K.}~\bibnamefont
  {Kumar}}, \bibinfo {author} {\bibfnamefont {N.~K.}\ \bibnamefont {Karn}}, \
  and\ \bibinfo {author} {\bibfnamefont {V.~P.~S.}\ \bibnamefont {Awana}},\
  }\href@noop {} {\enquote {\bibinfo {title} {Synthesis of possible room
  temperature superconductor lk-99:pb$_9$cu(po$_4$)$_6$o},}\ } (\bibinfo {year}
  {2023}{\natexlab{b}}),\ \Eprint {http://arxiv.org/abs/2307.16402}
  {arXiv:2307.16402 [cond-mat.supr-con]} \BibitemShut {NoStop}%
\bibitem [{\citenamefont {Krivovichev}(2023)}]{krivovichev2023crystal}%
  \BibitemOpen
  \bibfield  {author} {\bibinfo {author} {\bibfnamefont {S.~V.}\ \bibnamefont
  {Krivovichev}},\ }\href@noop {} {\enquote {\bibinfo {title} {The crystal
  structure of pb$_{10}$(po$_4$)$_6$o revisited: the evidence of
  superstructure},}\ } (\bibinfo {year} {2023}),\ \Eprint
  {http://arxiv.org/abs/2308.04915} {arXiv:2308.04915 [cond-mat.supr-con]}
  \BibitemShut {NoStop}%
\bibitem [{\citenamefont {Cabezas-Escares}\ \emph {et~al.}(2023)\citenamefont
  {Cabezas-Escares}, \citenamefont {Barrera}, \citenamefont {Cardenas},\ and\
  \citenamefont {Munoz}}]{cabezasescares2023theoretical}%
  \BibitemOpen
  \bibfield  {author} {\bibinfo {author} {\bibfnamefont {J.}~\bibnamefont
  {Cabezas-Escares}}, \bibinfo {author} {\bibfnamefont {N.~F.}\ \bibnamefont
  {Barrera}}, \bibinfo {author} {\bibfnamefont {C.}~\bibnamefont {Cardenas}}, \
  and\ \bibinfo {author} {\bibfnamefont {F.}~\bibnamefont {Munoz}},\
  }\href@noop {} {\enquote {\bibinfo {title} {Theoretical insight on the lk-99
  material},}\ } (\bibinfo {year} {2023}),\ \Eprint
  {http://arxiv.org/abs/2308.01135} {arXiv:2308.01135 [cond-mat.supr-con]}
  \BibitemShut {NoStop}%
\bibitem [{\citenamefont {Si}\ \emph {et~al.}(2023)\citenamefont {Si},
  \citenamefont {Wallerberger}, \citenamefont {Smolyanyuk}, \citenamefont
  {di~Cataldo}, \citenamefont {Tomczak},\ and\ \citenamefont
  {Held}}]{si2023pb10xcuxpo46o}%
  \BibitemOpen
  \bibfield  {author} {\bibinfo {author} {\bibfnamefont {L.}~\bibnamefont
  {Si}}, \bibinfo {author} {\bibfnamefont {M.}~\bibnamefont {Wallerberger}},
  \bibinfo {author} {\bibfnamefont {A.}~\bibnamefont {Smolyanyuk}}, \bibinfo
  {author} {\bibfnamefont {S.}~\bibnamefont {di~Cataldo}}, \bibinfo {author}
  {\bibfnamefont {J.~M.}\ \bibnamefont {Tomczak}}, \ and\ \bibinfo {author}
  {\bibfnamefont {K.}~\bibnamefont {Held}},\ }\href@noop {} {\enquote {\bibinfo
  {title} {$\mathrm{Pb_{10-x}Cu_x(PO_4)_6O}$: a mott or charge transfer
  insulator in need of further doping for (super)conductivity},}\ } (\bibinfo
  {year} {2023}),\ \Eprint {http://arxiv.org/abs/2308.04427} {arXiv:2308.04427
  [cond-mat.supr-con]} \BibitemShut {NoStop}%
\bibitem [{\citenamefont {Kresse}\ and\ \citenamefont
  {Hafner}(1993)}]{PhysRevB.47.558}%
  \BibitemOpen
  \bibfield  {author} {\bibinfo {author} {\bibfnamefont {G.}~\bibnamefont
  {Kresse}}\ and\ \bibinfo {author} {\bibfnamefont {J.}~\bibnamefont
  {Hafner}},\ }\href {\doibase 10.1103/PhysRevB.47.558} {\bibfield  {journal}
  {\bibinfo  {journal} {Phys. Rev. B}\ }\textbf {\bibinfo {volume} {47}},\
  \bibinfo {pages} {558} (\bibinfo {year} {1993})}\BibitemShut {NoStop}%
\bibitem [{\citenamefont {Kresse}\ and\ \citenamefont
  {Furthm\"uller}(1996)}]{PhysRevB.54.11169}%
  \BibitemOpen
  \bibfield  {author} {\bibinfo {author} {\bibfnamefont {G.}~\bibnamefont
  {Kresse}}\ and\ \bibinfo {author} {\bibfnamefont {J.}~\bibnamefont
  {Furthm\"uller}},\ }\href {\doibase 10.1103/PhysRevB.54.11169} {\bibfield
  {journal} {\bibinfo  {journal} {Phys. Rev. B}\ }\textbf {\bibinfo {volume}
  {54}},\ \bibinfo {pages} {11169} (\bibinfo {year} {1996})}\BibitemShut
  {NoStop}%
\bibitem [{\citenamefont {Dudarev}\ \emph {et~al.}(1998)\citenamefont
  {Dudarev}, \citenamefont {Botton}, \citenamefont {Savrasov}, \citenamefont
  {Humphreys},\ and\ \citenamefont {Sutton}}]{PhysRevB.57.1505}%
  \BibitemOpen
  \bibfield  {author} {\bibinfo {author} {\bibfnamefont {S.~L.}\ \bibnamefont
  {Dudarev}}, \bibinfo {author} {\bibfnamefont {G.~A.}\ \bibnamefont {Botton}},
  \bibinfo {author} {\bibfnamefont {S.~Y.}\ \bibnamefont {Savrasov}}, \bibinfo
  {author} {\bibfnamefont {C.~J.}\ \bibnamefont {Humphreys}}, \ and\ \bibinfo
  {author} {\bibfnamefont {A.~P.}\ \bibnamefont {Sutton}},\ }\href {\doibase
  10.1103/PhysRevB.57.1505} {\bibfield  {journal} {\bibinfo  {journal} {Phys.
  Rev. B}\ }\textbf {\bibinfo {volume} {57}},\ \bibinfo {pages} {1505}
  (\bibinfo {year} {1998})}\BibitemShut {NoStop}%
\bibitem [{\citenamefont {Liechtenstein}\ \emph {et~al.}(1995)\citenamefont
  {Liechtenstein}, \citenamefont {Anisimov},\ and\ \citenamefont
  {Zaanen}}]{PhysRevB.52.R5467}%
  \BibitemOpen
  \bibfield  {author} {\bibinfo {author} {\bibfnamefont {A.~I.}\ \bibnamefont
  {Liechtenstein}}, \bibinfo {author} {\bibfnamefont {V.~I.}\ \bibnamefont
  {Anisimov}}, \ and\ \bibinfo {author} {\bibfnamefont {J.}~\bibnamefont
  {Zaanen}},\ }\href {\doibase 10.1103/PhysRevB.52.R5467} {\bibfield  {journal}
  {\bibinfo  {journal} {Phys. Rev. B}\ }\textbf {\bibinfo {volume} {52}},\
  \bibinfo {pages} {R5467} (\bibinfo {year} {1995})}\BibitemShut {NoStop}%
\bibitem [{\citenamefont {Karolak}\ \emph {et~al.}(2011)\citenamefont
  {Karolak}, \citenamefont {Wehling}, \citenamefont {Lechermann},\ and\
  \citenamefont {Lichtenstein}}]{karolak_general_2011}%
  \BibitemOpen
  \bibfield  {author} {\bibinfo {author} {\bibfnamefont {M.}~\bibnamefont
  {Karolak}}, \bibinfo {author} {\bibfnamefont {T.~O.}\ \bibnamefont
  {Wehling}}, \bibinfo {author} {\bibfnamefont {F.}~\bibnamefont {Lechermann}},
  \ and\ \bibinfo {author} {\bibfnamefont {A.~I.}\ \bibnamefont
  {Lichtenstein}},\ }\href {\doibase 10.1088/0953-8984/23/8/085601} {\bibfield
  {journal} {\bibinfo  {journal} {Journal of Physics: Condensed Matter}\
  }\textbf {\bibinfo {volume} {23}},\ \bibinfo {pages} {085601} (\bibinfo
  {year} {2011})}\BibitemShut {NoStop}%
\bibitem [{\citenamefont {Schüler}\ \emph {et~al.}(2018)\citenamefont
  {Schüler}, \citenamefont {Peil}, \citenamefont {Kraberger}, \citenamefont
  {Pordzik}, \citenamefont {Marsman}, \citenamefont {Kresse}, \citenamefont
  {Wehling},\ and\ \citenamefont {Aichhorn}}]{S2018}%
  \BibitemOpen
  \bibfield  {author} {\bibinfo {author} {\bibfnamefont {M.}~\bibnamefont
  {Schüler}}, \bibinfo {author} {\bibfnamefont {O.~E.}\ \bibnamefont {Peil}},
  \bibinfo {author} {\bibfnamefont {G.~J.}\ \bibnamefont {Kraberger}}, \bibinfo
  {author} {\bibfnamefont {R.}~\bibnamefont {Pordzik}}, \bibinfo {author}
  {\bibfnamefont {M.}~\bibnamefont {Marsman}}, \bibinfo {author} {\bibfnamefont
  {G.}~\bibnamefont {Kresse}}, \bibinfo {author} {\bibfnamefont {T.~O.}\
  \bibnamefont {Wehling}}, \ and\ \bibinfo {author} {\bibfnamefont
  {M.}~\bibnamefont {Aichhorn}},\ }\href {\doibase 10.1088/1361-648X/aae80a}
  {\bibfield  {journal} {\bibinfo  {journal} {Journal of Physics: Condensed
  Matter}\ }\textbf {\bibinfo {volume} {30}},\ \bibinfo {pages} {475901}
  (\bibinfo {year} {2018})}\BibitemShut {NoStop}%
\bibitem [{\citenamefont {Amadon}\ \emph {et~al.}(2008)\citenamefont {Amadon},
  \citenamefont {Lechermann}, \citenamefont {Georges}, \citenamefont {Jollet},
  \citenamefont {Wehling},\ and\ \citenamefont {Lichtenstein}}]{amadon2008}%
  \BibitemOpen
  \bibfield  {author} {\bibinfo {author} {\bibfnamefont {B.}~\bibnamefont
  {Amadon}}, \bibinfo {author} {\bibfnamefont {F.}~\bibnamefont {Lechermann}},
  \bibinfo {author} {\bibfnamefont {A.}~\bibnamefont {Georges}}, \bibinfo
  {author} {\bibfnamefont {F.}~\bibnamefont {Jollet}}, \bibinfo {author}
  {\bibfnamefont {T.~O.}\ \bibnamefont {Wehling}}, \ and\ \bibinfo {author}
  {\bibfnamefont {A.~I.}\ \bibnamefont {Lichtenstein}},\ }\href {\doibase
  10.1103/PhysRevB.77.205112} {\bibfield  {journal} {\bibinfo  {journal} {Phys.
  Rev. B}\ }\textbf {\bibinfo {volume} {77}},\ \bibinfo {pages} {205112}
  (\bibinfo {year} {2008})}\BibitemShut {NoStop}%
\bibitem [{\citenamefont {Merker}\ and\ \citenamefont
  {Wondratschek}(1960)}]{Merker}%
  \BibitemOpen
  \bibfield  {author} {\bibinfo {author} {\bibfnamefont {L.}~\bibnamefont
  {Merker}}\ and\ \bibinfo {author} {\bibfnamefont {H.}~\bibnamefont
  {Wondratschek}},\ }\href {\doibase https://doi.org/10.1002/zaac.19603060105}
  {\bibfield  {journal} {\bibinfo  {journal} {Zeitschrift für anorganische und
  allgemeine Chemie}\ }\textbf {\bibinfo {volume} {306}},\ \bibinfo {pages}
  {25} (\bibinfo {year} {1960})},\ \Eprint
  {http://arxiv.org/abs/https://onlinelibrary.wiley.com/doi/pdf/10.1002/zaac.19603060105}
  {https://onlinelibrary.wiley.com/doi/pdf/10.1002/zaac.19603060105}
  \BibitemShut {NoStop}%
\bibitem [{\citenamefont {Werner}\ \emph {et~al.}(2015)\citenamefont {Werner},
  \citenamefont {Sakuma}, \citenamefont {Nilsson},\ and\ \citenamefont
  {Aryasetiawan}}]{PhysRevB.91.125142}%
  \BibitemOpen
  \bibfield  {author} {\bibinfo {author} {\bibfnamefont {P.}~\bibnamefont
  {Werner}}, \bibinfo {author} {\bibfnamefont {R.}~\bibnamefont {Sakuma}},
  \bibinfo {author} {\bibfnamefont {F.}~\bibnamefont {Nilsson}}, \ and\
  \bibinfo {author} {\bibfnamefont {F.}~\bibnamefont {Aryasetiawan}},\ }\href
  {\doibase 10.1103/PhysRevB.91.125142} {\bibfield  {journal} {\bibinfo
  {journal} {Phys. Rev. B}\ }\textbf {\bibinfo {volume} {91}},\ \bibinfo
  {pages} {125142} (\bibinfo {year} {2015})}\BibitemShut {NoStop}%
\bibitem [{\citenamefont {Imada}\ \emph {et~al.}(1998)\citenamefont {Imada},
  \citenamefont {Fujimori},\ and\ \citenamefont {Tokura}}]{mit}%
  \BibitemOpen
  \bibfield  {author} {\bibinfo {author} {\bibfnamefont {M.}~\bibnamefont
  {Imada}}, \bibinfo {author} {\bibfnamefont {A.}~\bibnamefont {Fujimori}}, \
  and\ \bibinfo {author} {\bibfnamefont {Y.}~\bibnamefont {Tokura}},\ }\href
  {\doibase 10.1103/RevModPhys.70.1039} {\bibfield  {journal} {\bibinfo
  {journal} {Rev. Mod. Phys.}\ }\textbf {\bibinfo {volume} {70}},\ \bibinfo
  {pages} {1039} (\bibinfo {year} {1998})}\BibitemShut {NoStop}%
\bibitem [{\citenamefont {Mott}(1949)}]{mott}%
  \BibitemOpen
  \bibfield  {author} {\bibinfo {author} {\bibfnamefont {N.~F.}\ \bibnamefont
  {Mott}},\ }\href {\doibase 10.1088/0370-1298/62/7/303} {\bibfield  {journal}
  {\bibinfo  {journal} {Proceedings of the Physical Society. Section A}\
  }\textbf {\bibinfo {volume} {62}},\ \bibinfo {pages} {416} (\bibinfo {year}
  {1949})}\BibitemShut {NoStop}%
\bibitem [{\citenamefont {Hubbard}\ and\ \citenamefont
  {Flowers}(1963)}]{hubbard}%
  \BibitemOpen
  \bibfield  {author} {\bibinfo {author} {\bibfnamefont {J.}~\bibnamefont
  {Hubbard}}\ and\ \bibinfo {author} {\bibfnamefont {B.~H.}\ \bibnamefont
  {Flowers}},\ }\href {\doibase 10.1098/rspa.1963.0204} {\bibfield  {journal}
  {\bibinfo  {journal} {Proceedings of the Royal Society of London. Series A.
  Mathematical and Physical Sciences}\ }\textbf {\bibinfo {volume} {276}},\
  \bibinfo {pages} {238} (\bibinfo {year} {1963})},\ \Eprint
  {http://arxiv.org/abs/https://royalsocietypublishing.org/doi/pdf/10.1098/rspa.1963.0204}
  {https://royalsocietypublishing.org/doi/pdf/10.1098/rspa.1963.0204}
  \BibitemShut {NoStop}%
\bibitem [{\citenamefont {Zaanen}\ \emph {et~al.}(1985)\citenamefont {Zaanen},
  \citenamefont {Sawatzky},\ and\ \citenamefont {Allen}}]{ct}%
  \BibitemOpen
  \bibfield  {author} {\bibinfo {author} {\bibfnamefont {J.}~\bibnamefont
  {Zaanen}}, \bibinfo {author} {\bibfnamefont {G.~A.}\ \bibnamefont
  {Sawatzky}}, \ and\ \bibinfo {author} {\bibfnamefont {J.~W.}\ \bibnamefont
  {Allen}},\ }\href {\doibase 10.1103/PhysRevLett.55.418} {\bibfield  {journal}
  {\bibinfo  {journal} {Phys. Rev. Lett.}\ }\textbf {\bibinfo {volume} {55}},\
  \bibinfo {pages} {418} (\bibinfo {year} {1985})}\BibitemShut {NoStop}%
\bibitem [{\citenamefont {Jain}(2023)}]{jain2023phase}%
  \BibitemOpen
  \bibfield  {author} {\bibinfo {author} {\bibfnamefont {P.~K.}\ \bibnamefont
  {Jain}},\ }\href@noop {} {\enquote {\bibinfo {title} {Phase transition of
  copper (i) sulfide and its implication for purported superconductivity of
  lk-99},}\ } (\bibinfo {year} {2023}),\ \Eprint
  {http://arxiv.org/abs/2308.05222} {arXiv:2308.05222 [cond-mat.supr-con]}
  \BibitemShut {NoStop}%
\bibitem [{\citenamefont {Thakur}\ \emph {et~al.}(2023)\citenamefont {Thakur},
  \citenamefont {Schulze},\ and\ \citenamefont {Ruck}}]{thakur2023synthesis}%
  \BibitemOpen
  \bibfield  {author} {\bibinfo {author} {\bibfnamefont {G.~S.}\ \bibnamefont
  {Thakur}}, \bibinfo {author} {\bibfnamefont {M.}~\bibnamefont {Schulze}}, \
  and\ \bibinfo {author} {\bibfnamefont {M.}~\bibnamefont {Ruck}},\ }\href@noop
  {} {\enquote {\bibinfo {title} {On the synthesis methodologies to prepare
  pb$_9$cu(po$_4$)$_6$o -- phase, composition, magnetic analysis and absence of
  superconductivity},}\ } (\bibinfo {year} {2023}),\ \Eprint
  {http://arxiv.org/abs/2308.05776} {arXiv:2308.05776 [cond-mat.supr-con]}
  \BibitemShut {NoStop}%
\end{thebibliography}%

\end{document}